# An Application for Research – the Large Hadron Collider

*R. Bailey*
CERN, Geneva, Switzerland

**Abstract**
The Large Hadron Collider (LHC) machine at CERN was designed and built primarily to find or exclude the existence of the Higgs boson, for which a large amount of data is needed by the LHC experiments. This requires operation at high luminosity, which in turn requires running with thousands of high-intensity proton bunches in the machine. After quantifying the data required by the experiments and elucidating the LHC parameters needed to achieve this, this paper explains how the LHC beams are fabricated from the pulse(s) coming from the CERN Duoplasmatron source.

## 1   Introduction

At the start of 2012, data from Large Electron–Positron collider (LEP), the Tevatron and the Large Hadron Collider (LHC) excluded the existence of the Higgs boson for the mass regions $m_\text{H} < 114$ GeV and $m_\text{H} > 150$ GeV. The expected cross-section for the production of the Higgs boson in the remaining mass range at LHC energies is measured in tens of picobarns (pb), to be compared with the total inelastic proton–proton cross-section, measured in tens of millibarns (mb) (1 barn = $10^{-28}$ m$^2$). When combined with the expected decay branching ratios to the experimentally most sensitive channels at LHC, values of the quantity cross-section × branching ratio are measured in femtobarns (fb). Hence large amounts of data are needed to produce and detect merely a handful of Higgs boson events.

The quantity used to define the instantaneous performance level of a collider is luminosity, $L$, measured in cm$^{-2}$ s$^{-1}$. An average luminosity $\langle L \rangle$ delivered over a period of time $t$ in collision yields a value of the delivered luminosity $L_\text{DEL} = \langle L \rangle t$, measured in barn$^{-1}$. Running for $10^6$ s at an average luminosity of $10^{33}$ cm$^{-2}$ s$^{-1}$ yields an integrated luminosity of 1 fb$^{-1}$.

LHC therefore has to operate routinely at high luminosity. The design value is $10^{34}$ cm$^{-2}$ s$^{-1}$, which requires a number of parameters to be achieved, not least the huge number of bunches in each counter-rotating beam. This is realized by a complex series of manipulations in the LHC injector chain.

## 2   Cross-sections and branching ratios

The expected cross-sections for production of the Higgs boson at $\sqrt{s}$ of 7 TeV (LHC in 2011) are shown in Fig. 1 [1]. For $m_\text{H} < 1$ TeV, the dominant process is gluon fusion and the expected cross-section is of the order of 10 pb in the Higgs mass range not already excluded.

The expected branching ratios for the decay of the Higgs, as a function of the Higgs mass, are shown in Fig. 2 [1]. For example, at a Higgs mass ~125 GeV, the branching ratio Higgs → ZZ is around $2 \times 10^{-2}$. To detect the Z through its decay into any of the three lepton families, a further branching ratio of $10^{-1}$ has to be applied to the decay of each Z. Therefore, the resulting branching ratio of Higgs → ZZ → four leptons is around $2 \times 10^{-4}$.

For a Higgs mass of 125 GeV, combining the production cross-section ($\sigma$) at $\sqrt{s}$ of 7 TeV with the branching ratio (BR) of Higgs → ZZ → four leptons, we obtain a value of $\sigma$BR of 2 fb.

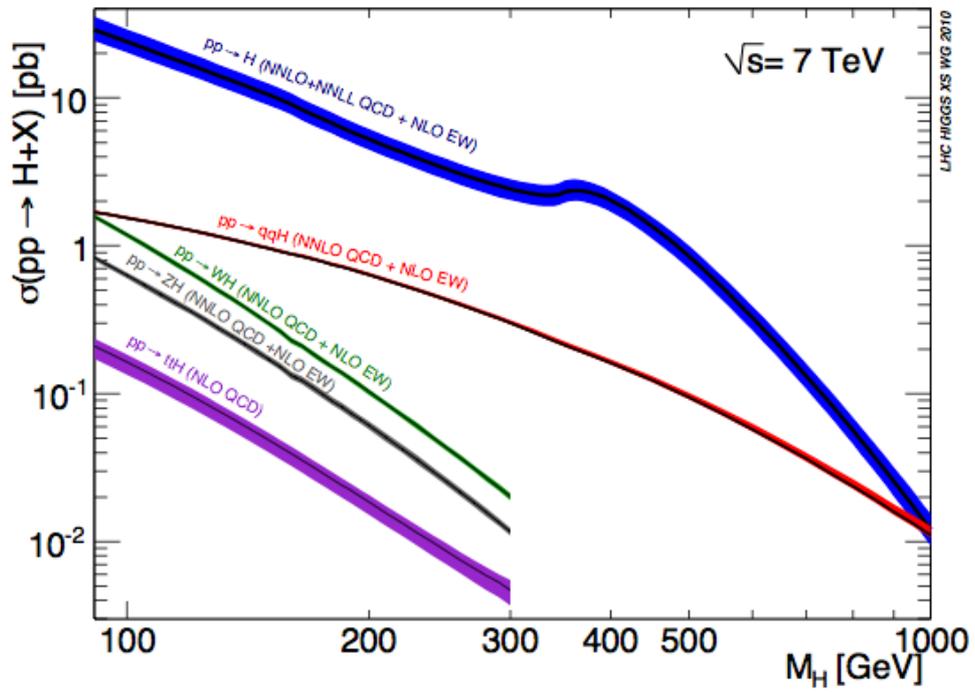

**Fig. 1:** Expected cross-sections for production of the Higgs boson at $\sqrt{s}$ = 7 TeV

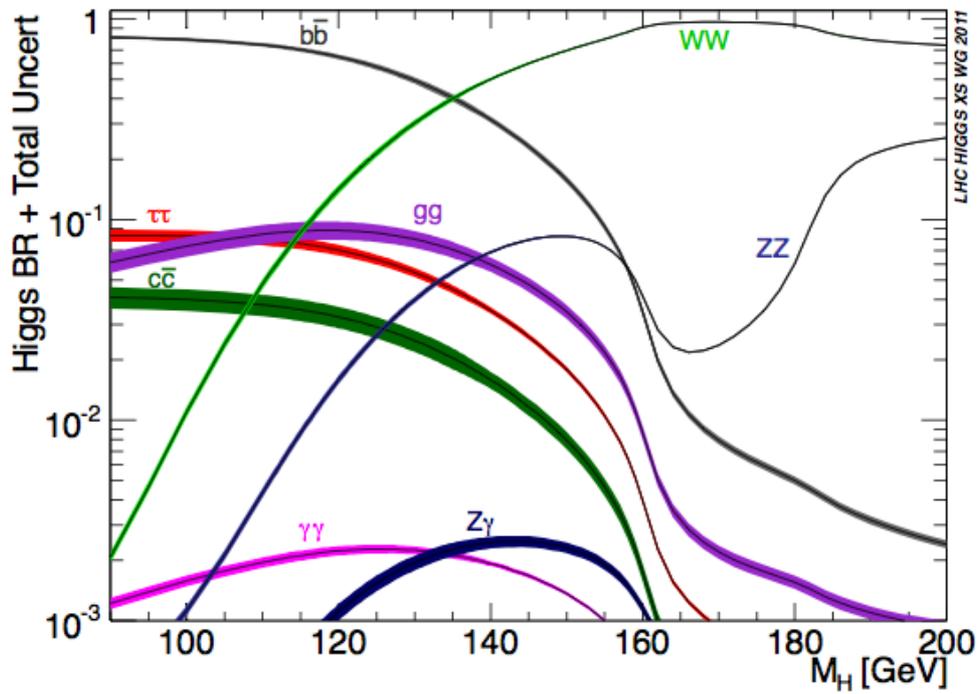

**Fig. 2:** Expected Higgs decay branching ratios

## 3 Luminosity and associated parameters

The performance of a collider is characterized by the luminosity delivered, $L_{DEL}$, over a period of time. If the average instantaneous luminosity over the time period $t$ is $\langle L \rangle$, then

$$L_{DEL} = \langle L \rangle t \text{ barn}^{-1}. \quad (1)$$

The number of events $N_{event}$ with production cross-section $\sigma$ and branching ratio BR detected in these data is given by

$$N_{event} = \sigma \, BR \, L_{DEL}. \quad (2)$$

For the Higgs → ZZ → four leptons process mentioned above, a delivered luminosity of 5 fb$^{-1}$ would yield just 10 events. In 2011, the LHC provided colliding beams for approximately $5 \times 10^6$ s in order to deliver 5 fb$^{-1}$ of integrated luminosity, which from Eq. (1) corresponds to an average instantaneous luminosity of $10^{33}$ cm$^{-2}$ s$^{-1}$. When one considers that the luminosity decays during a physics fill of several hours' duration, it is clear that instantaneous luminosities well in excess of $10^{33}$ are required to deliver even tens of these kinds of events per year at the LHC.

The luminosity at any moment can be written as

$$L = \frac{N^2 k_b f \gamma}{4\pi \varepsilon_n \beta^*} F \quad (3)$$

where $\gamma$, the relativistic factor, is given by the beam energy $E$ divided by the proton rest mass, $f$ is the revolution frequency, and $F$ is a reduction factor resulting from operating with a crossing angle, $\theta_c$.

The explanation of the other variables, together with LHC parameters for nominal, 2011 values and aims for 2012, is given in Table 1.

**Table 1:** LHC parameters in 2011 and 2012 compared to nominal.

| Symbol | Meaning | Units | Nominal | 2011 | 2012 |
|---|---|---|---|---|---|
| $E$ | Beam energy | TeV | 7.0 | 3.5 | 4.0 |
| $N$ | Number of particles per bunch | | $1.15 \times 10^{11}$ | $1.5 \times 10^{11}$ | $1.5 \times 10^{11}$ |
| $k_b$ | Number of bunches per beam† | | 2808 | 1380 | 1380 |
| $\varepsilon_n$ | Normalized emittance ($= \varepsilon\gamma$) | μm rad | 3.75 | 2.5 | 2.5 |
| $\beta^*$ | Beat function at IP 1 and 5 | m | 0.55 | 1.0 | 0.6 |
| $\theta_c$ | Crossing angle through IP 1 and 5 | μrad | 285 | 240 | 290 |
| $L$ | **Peak instantaneous luminosity** | cm$^{-2}$ s$^{-1}$ | $10^{34}$ | $3.5 \times 10^{33}$ | $6 \times 10^{33}$ |

†Owing to the details of the filling scheme, a few per cent of bunches do not contribute to luminosity.

While the energy, the beta function at the interaction point (IP) and the crossing angle are all very much in the domain of the LHC itself, the collider is reliant on the injector chain to provide beams with thousands of high-intensity bunches of small transverse emittance.

# 4 The LHC injector chain

The LHC is supplied with protons from an injector chain comprising accelerators that have been in use for many years, namely Linac2, the Proton Synchrotron Booster (PSB), the Proton Synchrotron (PS) and the Super Proton Synchrotron (SPS), as shown in Fig. 3. These accelerators were all upgraded in the 1990s to meet the very stringent needs of the LHC.

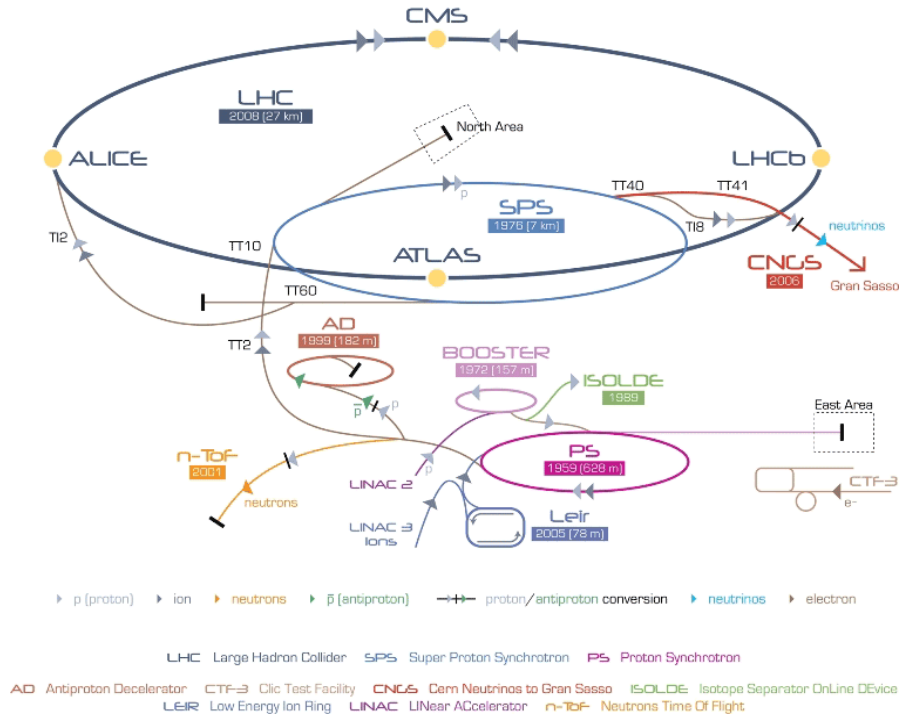

**Fig. 3:** The CERN accelerator complex

Linac2 has been the primary source of protons for the CERN accelerator complex for the past 30 years and has been modified to meet the demands of the LHC. The ion source is a Duoplasmatron giving up to 300 mA of beam current, followed by a four-vane Radio Frequency Quadrupole (RFQ) with output energy of 750 keV. A three-tank drift-tube linac with quadrupole focusing in the drift tubes follows. An 80 m long transfer line then carries the beam at 50 MeV to the entrance to the PSB.

The PSB is actually four rings stacked on top of each other. Each has one-quarter of the PS circumference, allowing beam to be extracted sequentially from each ring towards the PS, and thus in principle filling the circumference of the PS machine. The extraction energy of the PSB was increased from 1 GeV to 1.4 GeV in order to decrease the space-charge problems at injection in the PS machine.

The PS accelerates the beam to 14 GeV (for SPS physics beams) or to 26 GeV (for LHC beams) before transfer to the SPS, which in turn can accelerate the beam up to an energy of 450 GeV. The circumference ratios of PS/SPS and SPS/LHC are 1/11 and 7/27, respectively.

# 5 Production of LHC bunch trains

The LHC radiofrequency (RF) system operates on $h = 35\,640$ at a frequency of 400 MHz, which with one in ten buckets used for beam yields 3564 bunch places spaced by 25 ns to completely fill the circumference. In practice, not all of the bunch places can be used in order to leave gaps for the various kickers used in the injection and extraction processes. The optimum scheme developed [2] results in 2808 bunches in each ring of the LHC for nominal performance.

## 5.1 From PS extraction to LHC

The 2808 bunches in each LHC ring are provided by multiple injections from the SPS, which in turn is fed by multiple injections from the PS. For each extraction or injection, a fast kicker magnet is used, and the characteristics of these kickers determine the minimum gaps needed in the trains of bunches at various stages of beam production.

Figure 4 summarizes the bunch trains from PS extraction to LHC injection. A PS batch consists of 72 bunches on $h = 84$ at extraction. Either three or four of these batches are sequentially transferred to the SPS, thereby partially filling 3/11 or 4/11 of the SPS circumference. For each LHC ring, 12 of these 216 or 288 bunch trains are transferred from SPS to LHC. With 9×216 + 3×288 injections, the LHC is filled with 2808 bunches.

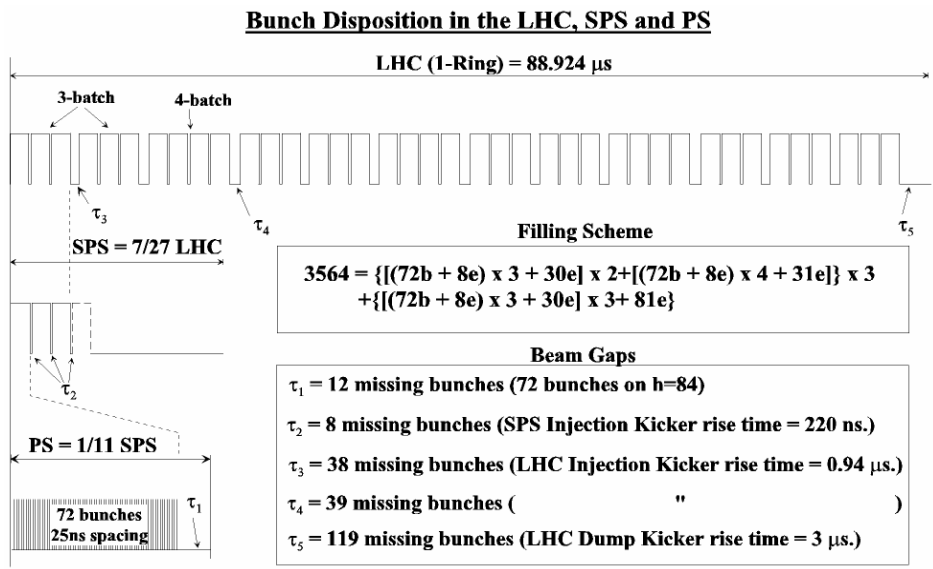

**Fig. 4:** Bunches in the LHC, SPS and PS machines

The SPS cycle needed to allow up to four injections from the PS, followed by the acceleration from 25 GeV to 450 GeV, followed by extraction to the LHC, has a duration of 21.6 s. Therefore, in an ideal situation, where every transfer works perfectly, the time needed to fill each LHC ring is a little under 5 min. In practice, this takes significantly longer.

## 5.2 From PSB to PS injection

With the PSB operating on $h = 1$, one bunch per ring is transferred to the PS, operating on $h = 7$. It is possible to fill the 6/7 of the PS ring with six bunches using a double-batch filling scheme, with either four bunches in the first batch and two in the second, or with three bunches per batch as depicted in Fig. 5. Each PSB cycle is 1.2 s long.

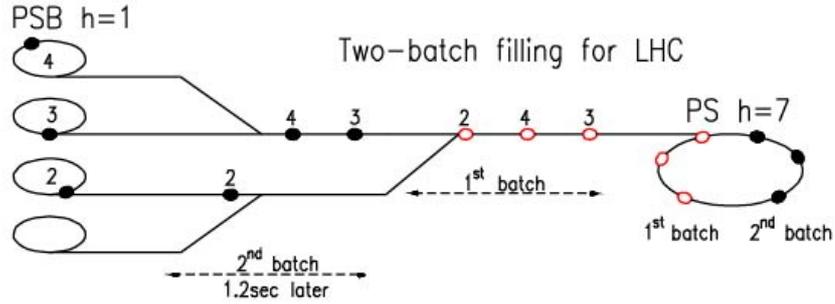

**Fig. 5:** PS double-batch filling scheme for the LHC

### 5.3 Bunch splitting in the PS

We have seen that the standard LHC filling scheme is based on a PS batch of 72 bunches being sent to the SPS, while the PSB fills the PS circumference with only six bunches. The six become 72 through a multiple splitting scheme in the PS machine [3].

Bunches from the PSB arrive in the PS and are captured by one of the PS RF systems. This is illustrated in Fig. 6, where the use of a sinusoidal RF waveform provides a stationary bucket in longitudinal phase space, into which the injected bunch arrives (provided that the phase is correct).

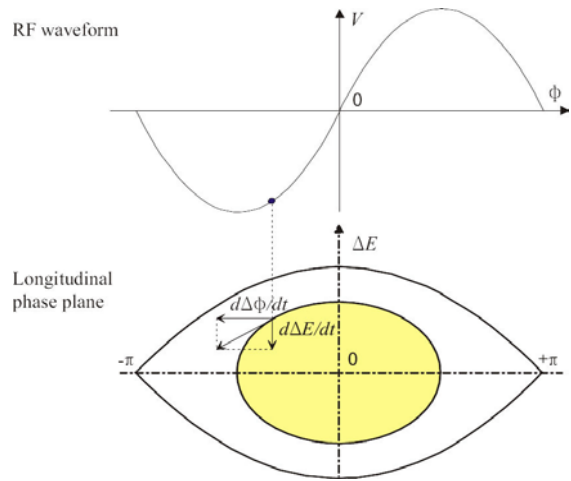

**Fig. 6:** Capture into a stationary bucket

The principle of using more than one RF waveform is shown in Fig. 7. By the simultaneous application of two RF waveforms, on $h$ and $2h$, and an adiabatic change of the voltages as shown, the single stationary bucket can be transformed into two stationary buckets in the higher-frequency RF system. The bunch that was captured in the lower-frequency system is split into two bunches in the higher-frequency system.

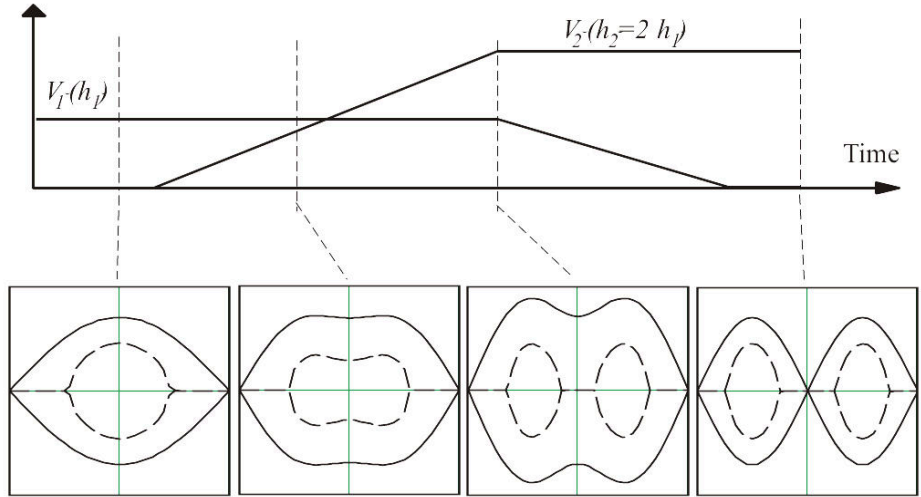

**Fig. 7:** Bunch splitting principle

Adding a third RF waveform, with appropriate amplitude and phase parameters, allows a bunch to be split into three. This is shown in simulation in Fig. 8(a), with real data shown in Fig. 8(b).

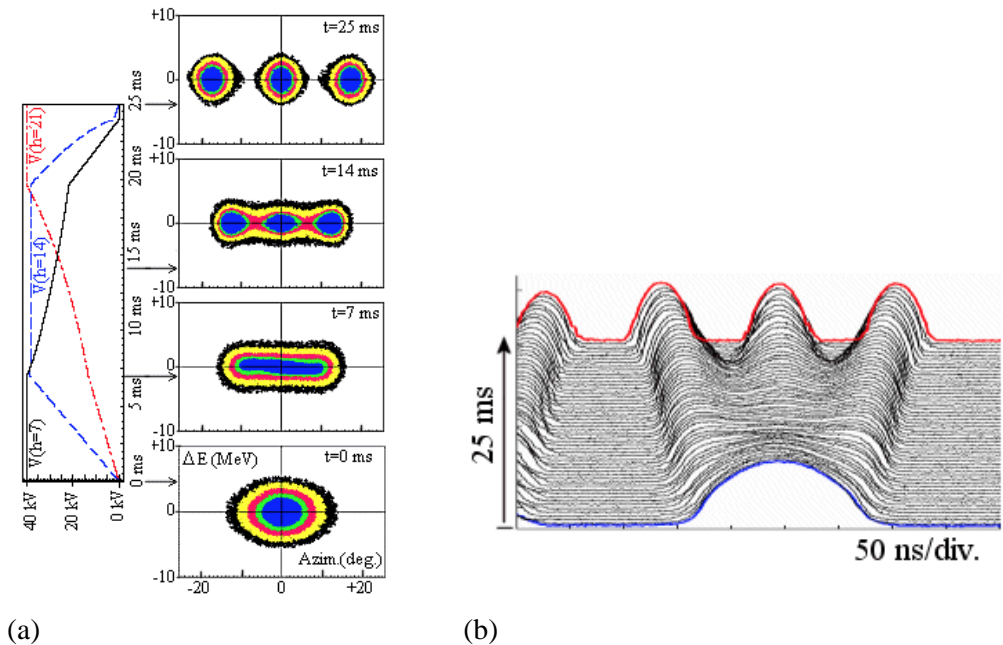

(a)          (b)

**Fig. 8:** Triple bunch splitting in the PS

These techniques were developed and perfected in the 1990s and 2000s in the PS for the production of the LHC beam [4]. The complete process presently used is shown in Fig. 9.

Six PSB bunches are captured on harmonic 7 in the PS. The bunches are then split in three on the 1.4 GeV injection plateau using three groups of RF cavities operating on harmonics 7, 14 and 21. When bunched on $h = 21$, the beam is accelerated to 25 GeV, where each bunch is split twice (quadruple splitting) by consecutive application of RF systems operating on harmonics 42 and 84, at 20 MHz and 40 MHz, respectively. Finally, an 80 MHz system is used to perform bunch rotation, in order to shorten each bunch to a length of 4 ns before extraction, in order to fit into the buckets of the SPS 200 MHz RF system.

Note the development of the empty bucket (only six bunches are captured by the $h = 7$ system) to leave a beam gap allowing the clean voltage rise of the PS extraction kicker.

Thus six bunches injected become 72 bunches extracted, with bunch length 4 ns, through the careful manipulation of the voltage waveforms of six RF systems in the PS machine.

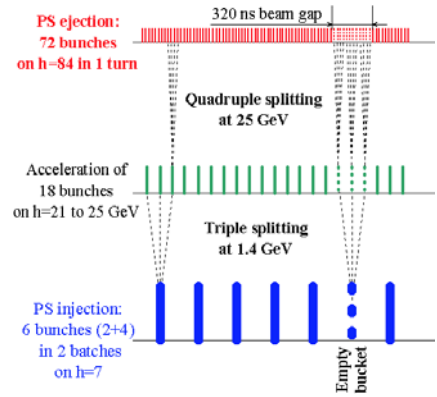

**Fig. 9:** Bunch evolution in the PS

## 5.4 Intensities

For nominal LHC operation, the single bunch intensity is $1.15 \times 10^{11}$ protons. With the scheme described, the implications for the different machines are summarized in Table 2, assuming zero losses throughout the entire chain. This is too idealistic. Each injection process, each acceleration cycle and each extraction process are never loss-free, and with four machines involved the cumulative losses become significant. Taking a combined loss of 10% for each extraction or injection process, and a 10% loss through the acceleration cycle, yields the intensities shown in Table 3. Comparing the booster intensity in the two cases shows approximately a factor of two arising from the cumulative losses.

**Table 2:** Intensities in the various machines for nominal LHC beams, no losses.

| Machine | Bunch intensity | Number of bunches | Total intensity | Scheme |
|---|---|---|---|---|
| LHC | $1.15 \times 10^{11}$ | 2808 | $3.23 \times 10^{14}$ | 12 SPS cycles, 234 334 334 334 |
| SPS | $1.15 \times 10^{11}$ | 288 | $3.31 \times 10^{13}$ | Four PS batches of 72 |
| PS | $1.15 \times 10^{11}$ | 72 | $8.28 \times 10^{12}$ | Six injections, each split ×12 |
| PSB | $1.38 \times 10^{12}$ | 1 | $1.38 \times 10^{12}$ | Two-turn injection into PSB |

Table 3: Intensities in the various machines for nominal LHC beams, 10% losses.

| Machine | Bunch intensity | Number of bunches | Total intensity | Scheme |
|---|---|---|---|---|
| LHC | $1.15 \times 10^{11}$ | 2808 | $3.55 \times 10^{14}$ | 12 SPS cycles, 234 334 334 334 |
| SPS | $1.39 \times 10^{11}$ | 288 | $4.41 \times 10^{13}$ | Four PS batches of 72 |
| PS | $1.68 \times 10^{11}$ | 72 | $1.33 \times 10^{13}$ | Six injections, each split ×12 |
| PSB | $2.22 \times 10^{12}$ | 1 | $2.44 \times 10^{12}$ | Two-turn injection into PSB |

## 6   Requirements on the Linac2 complex

In order to fill each ring of the PSB, Linac2 has to deliver in a single pulse, in the most demanding case, four times the PSB intensity, or around $10^{13}$ protons. This is the intensity needed at the entrance to the PSB. Losses in the various parts of the Linac2 complex (50 MeV transfer, RFQ, source) push the required intensity from the source to somewhere closer to $2 \times 10^{13}$ protons.

The PSB revolution time at the proton injection energy of 50 MeV is 1.67 µs, and since each PSB ring is filled using a two-turn low-emittance injection scheme, the pulse length of the beam coming from the linac/source has to be at least 13 µs.

Typical operating conditions therefore required for nominal LHC beams are for a current during the pulse of 165 mA in a 20 µs pulse, which corresponds to around $2 \times 10^{13}$ protons. In practice, these requirements have been surpassed, enabling the Linac2 complex to provide higher bunch intensities than nominal.

## 7   Transverse emittance

For nominal performance of the LHC, the normalized emittance $\varepsilon_n$ of the colliding bunches should be kept below 3.75 µm rad (Table 1).

Each bunch of protons emanating from the source is:

(i)   accelerated in Linac2 and transferred to the PSB;

(ii)  injected into (one ring of) the PSB, accelerated, extracted and transferred to the PS;

(iii) injected into the PS, split, accelerated, split, extracted and transferred to the SPS;

(iv)  injected into the SPS, accelerated, extracted and transferred to the LHC; and

(v)   injected into the LHC, accelerated, squeezed and brought into collision.

Each of these many operations (and more not mentioned) could lead to blow-up of the transverse emittance. In order to meet the LHC requirement, the emittance budget in the LHC Design Report specifies emittances for nominal bunches at extraction from the PBS, PS and SPS machines to be 2.5 µm rad, 3 µm rad and 3.5 µm rad, respectively.

In practice, the performance of the injectors has far exceeded these specifications. In 2011 the average values for normalized emittances in the three machines, corresponding to bunch populations in the SPS of $1.5 \times 10^{11}$, were 1.2 µm rad, 1.6 µm rad and 1.8 µm rad [5]. This has allowed colliding beam emittances in the LHC of under 2.5 µm rad, making a considerable contribution to the luminosity delivered. It should be noted that these figures correspond to operation with 50 ns bunch spacing, where one less bunch splitting is needed in the PS. This could contribute to the lower emittance.

## 8    Summary

The LHC injector chain plays a crucial role in the performance of the LHC. The fabrication of the thousands of bunches needed for high-luminosity performance involves a complex series of manipulations, each of which has to be made without significant detrimental impact on the transverse emittance of the beam being provided to the LHC. This not only has to be realized, but also has to be maintained throughout several months of operation per year.

The fact that LHC delivered luminosity levels that allowed the discovery of the Higgs boson in 2012 is testimony to the ingenuity of the physicists and engineers who have developed the injection scheme, and to the dedicated professionalism of the operations personnel who ensured that high performance was regularly delivered.